\documentclass[twoside]{book}
\usepackage{amsmath}
\usepackage{amsfonts}
\usepackage{amssymb}
\usepackage{graphicx}		

 \makeatletter
\ifx\UNDEF\mail\def\mail{ }\else\fi
\ifx\UNDEF\prange\def\prange{0 0}\else\fi

\gdef\@empty{}
\def\Mail#1 #2 {\gdef\thecontact{#1}\gdef\theaddr{#2}}
\def\Range#1 #2 {\gdef\thefirstpage{#1}\gdef\thelastpage{#2}}
{\let\'\mail \expandafter\Mail\' }  
{\let\'\prange \expandafter\Range\' }   
 \gdef\@shtitle{\relax}
 \long\def\shtitle#1{\gdef\@shtitle{#1}}
 \long\def\author#1{\gdef\@author{#1}}
 \def\affil#1{\par\noindent{\rm#1\par}}
 \gdef\@abstract{}
 \long\def\abstract#1{\gdef\@abstract{#1}}

 \def\maketitle{\thispagestyle{empty}\chapter{\@title}}
 \renewcommand\chapter{\if@openright\cleardoublepage\else\clearpage\fi
                    \thispagestyle{empty}%
                    \global\@topnum\z@
                    \@afterindentfalse
                    \secdef\@chapter\@schapter}
 \def\@makechapterhead#1{%
  \vspace*{50\p@}%
  {\parindent \z@ \raggedleft \normalfont
    \ifnum \c@secnumdepth >\m@ne
      \if@mainmatter
        \par\nobreak
        \vskip 20\p@
      \fi
    \fi
    \interlinepenalty\@M
    \Huge \bfseries #1\par\nobreak
    \vskip.25in
    \large\bfseries\@author\par\nobreak
    \vskip 40\p@}
    \ifx\@abstract\@empty\else{\small\@abstract\par\vskip20\p@}\fi
  }

\DeclareRobustCommand\em
        {\@nomath\em \ifdim \fontdimen\@ne\font >\z@
                       \upshape \else \slshape \fi}

\def\@begintheorem#1#2{\sl \trivlist \item[\hskip \labelsep{\bf #1\ #2}]}
\def\@opargbegintheorem#1#2#3{\sl \trivlist
     \item[\hskip \labelsep{\bf #1\ #2\ (#3)}]}

  \def\@arabic#1{\number #1} 

\long\def\@makecaption#1#2{
    \vskip\abovecaptionskip
    \sbox\@tempboxa{{\small {\bf #1}: #2}}%
    \ifdim\wd\@tempboxa>\hsize
        {\small {\bf #1}: #2\par}
    \else
       \global\@minipagefalse
       \hbox to\hsize{\hfil\box\@tempboxa\hfil}
    \fi
    \vskip \belowcaptionskip}

\def\figstrut#1{\hbox to\linewidth{\vrule height#1\hfill}}

\renewenvironment{thebibliography}[1]
     {\section*{\bibname
        \@mkboth{\MakeUppercase\bibname}{\MakeUppercase\bibname}}%
      \list{\@biblabel{\@arabic\c@enumiv}}%
           {\settowidth\labelwidth{\@biblabel{#1}}%
            \leftmargin\labelwidth
            \advance\leftmargin\labelsep
            \@openbib@code
            \usecounter{enumiv}%
            \let\p@enumiv\@empty
            \renewcommand\theenumiv{\@arabic\c@enumiv}}%
      \sloppy
      \clubpenalty4000
      \@clubpenalty \clubpenalty
      \widowpenalty4000%
      \sfcode`\.\@m}
     {\def\@noitemerr
       {\@latex@warning{Empty `thebibliography' environment}}%
      \endlist}
\makeatother

 \shtitle{Updating Probabilities}	
 \title{Updating Probabilities: A Complex Agent Based Example}
 \author{Adom Giffin\affil{Department of Physics\\ University at Albany--SUNY\\ Albany, NY 12222,USA}}
 \abstract{It has been shown that one can accommodate data (Bayes) and constraints (MaxEnt) in one method, the method of Maximum (relative) Entropy (ME) (Giffin 2007). In this paper we show a complex agent based example of inference with two different forms of information; moments and data. In this example, several agents each receive partial information about a system in the form of data. In addition, each agent agrees or is informed that there are certain global constraints on the system that are always true. The agents are then asked to make inferences about the entire system. The system becomes more complex as we add agents and allow them to share information. This system can have a geometrical form, such as a crystal structure. The shape may dictate how the agents are able to share information, such as sharing with nearest neighbors. This method can be used to model many systems where the agents or cells have local or partial information but must adhere to some global rules.}

\begin{document}           
\maketitle

\section{Introduction}

\label{intro}

There are many examples of systems where agents respond to both local
information as well as global information. Nature yields many such examples
where cells react to local stimuli yet carry some global instructions, such
as reproduction. The examples get more complex when the cells interact
locally or share information. This is the case in physics when one has a
lattice or group of many atoms where each is only affected by its nearest neighbor.
In all of these cases we would like to infer something about the system or
better, what each \emph{agent} infers about the system. It is this latter
case that we will be specifically addressing. The main purpose of this paper
is to examine a situation where each agent in a network (of varying degrees
of complexity) infers something about the whole system based on limited
information. By doing this we hope to attain clues about the system's
emergent properties, such as its dynamics, evolution, etc.

The two preeminent inference methods are the MaxEnt \cite{Jaynes57} method,
which has evolved to a more general method, the method of Maximum (relative)
Entropy (ME) \cite{ShoreJohnson80,Skilling88,CatichaGiffin06} and Bayes'
rule. The choice between the two methods has traditionally been dictated by
the nature of the information being processed (either constraints or
observed data). However, it has been shown that one can accommodate both
types of information in one method, ME \cite{GiffinCaticha07}. In fact, this
new ME method can reproduce every aspect of Bayesian and MaxEnt inference 
\emph{and} tackle problems that the two methods alone could not address. In
this paper we will show how the ME method can be used to infer properties of
the system under investigation.

We start by showing a general example of the ME method by inferring a
probability with two different forms of information: expected values%
\footnote{%
For simplicity we will refer to these expected values as \emph{moments}
although they can be considerably more general.} and data, \emph{%
simultaneously}. The solution resembles Bayes' Rule. In fact, if there are
no moment constraints then the method produces Bayes rule \emph{exactly}. If
there is no data, then the MaxEnt solution is produced.

Finally we solve a toy problem where we include global information in the
form of a moment constraint or expected value and then introduce local
information in the form of data. This will show how the agents infer aspects
of the whole system using the same process yet come to different
conclusions. Complexity is increased as the number of agents are increased
yet the complexity of the process does not grow proportionately. This
illustrates the advantages to using the ME method.

\section{Simultaneous updating}

\label{sec:2}

Our first concern when using the ME method to update from a prior to a
posterior distribution\footnote{%
In Bayesian inference, it is assumed that one always has a prior probability
based on some prior information. When new information is attained, the old
probility (the prior) is \emph{updated} to a new probability (the
posterior). If one has no prior information, then one uses an \emph{ignorant}
prior \cite{Gelman04}.} is to define the space in which the search for the
posterior will be conducted. We wish to infer something about the values of
one or several quantities, $\theta \in \Theta $, on the basis of three
pieces of information: prior information about $\theta $ (the prior), the
known relationship between $x$ \emph{and} $\theta $ (the model), and the
observed values of the data $x\in \mathcal{X}$. Since we are concerned with
both $x$ \emph{and} $\theta $, the relevant space is neither $\mathcal{X}$
nor $\Theta $ but the product $\mathcal{X}\times \Theta $ and our attention
must be focused on the joint distribution $P(x,\theta )$. The selected joint
posterior $P_{\text{new}}(x,\theta )$ is that which maximizes the entropy,%
\begin{equation}
S[P,P_{\text{old}}]=-\int dxd\theta ~P(x,\theta )\log \frac{P(x,\theta )}{P_{%
\text{old}}(x,\theta )}~,  \label{entropy}
\end{equation}%
subject to the appropriate constraints. $P_{\text{old}}(x,\theta )$ contains
our prior information which we call the \emph{joint prior}. To be explicit,%
\begin{equation}
P_{\text{old}}(x,\theta )=P_{\text{old}}(\theta )P_{\text{old}}(x|\theta )~,
\label{joint prior}
\end{equation}%
where $P_{\text{old}}(\theta )$ is the traditional Bayesian prior and $P_{%
\text{old}}(x|\theta )$ is the likelihood. It is important to note that they 
\emph{both} contain prior information. The Bayesian prior is defined as
containing prior information. However, the likelihood is not traditionally
thought of in terms of prior information. Of course it is reasonable to see
it as such because the likelihood represents the model (the relationship
between $\theta $ and $x)$ that has already been established. Thus we
consider both pieces, the Bayesian prior and the likelihood to be \emph{prior%
} information.

The new information is the \emph{observed data}, $x^{\prime }$, which in the
ME framework must be expressed in the form of a constraint on the allowed
posteriors. The family of posteriors that reflects the fact that $x$ is now
known to be $x^{\prime }$ is such that%
\begin{equation}
C_{1}:P(x)=\int d\theta ~P(x,\theta )=\delta (x-x^{\prime })~.  \label{data}
\end{equation}%
This amounts to an \emph{infinite} number of constraints: there is one
constraint on $P(x,\theta )$ for each value of the variable $x$ and each
constraint will require its own Lagrange multiplier $\lambda (x)$.
Furthermore, we impose the usual normalization constraint, 
\begin{equation}
\int dxd\theta ~P(x,\theta )=1~,  \label{Normalization}
\end{equation}%
and include additional information about $\theta $ in the form of a
constraint on the expected value of some function $f(\theta )$\footnote{%
Including an additional constraint in the form of $\int dxd\theta P(x,\theta
)g(x)=\left\langle g\right\rangle =G$ could only be used when it does not
contradict the data constraint (\ref{data}). Therefore, it is redundant and
the constraint would simply get absorbed when solving for $\lambda (x)$.}, 
\begin{equation}
C_{2}:\int dxd\theta \,P(x,\theta )f(\theta )=\left\langle f(\theta
)\right\rangle =F~.  \label{moment}
\end{equation}%
We emphasize that constraints imposed at the level of the prior need not be
satisfied by the posterior. What we do here differs from the standard
Bayesian practice in that we \emph{require} the constraint to be satisfied
by the posterior distribution.

Maximize (\ref{entropy}) subject to the above constraints, 
\begin{equation}
\delta \left\{ 
\begin{array}{c}
S+\alpha \left[ \int dxd\theta P(x,\theta )-1\right] \\ 
+\beta \left[ \int dxd\theta P(x,\theta )f(\theta )-F\right] \\ 
+\int dx\lambda (x)\left[ \int d\theta P(x,\theta )-\delta (x-x%
{\acute{}}%
)\right]%
\end{array}%
\right\} =0~,  \label{max}
\end{equation}%
yields the joint posterior,%
\begin{equation}
P_{\text{new}}(x,\theta )=P_{\text{old}}(x,\theta )\frac{e^{\lambda
(x)+\beta f(\theta )}}{Z}~,
\end{equation}%
where $Z$ is determined by using (\ref{Normalization}),%
\begin{equation}
Z=e^{-\alpha +1}=\int dxd\theta e^{\lambda (x)+\beta f(\theta )}P_{\text{old}%
}(x,\theta )
\end{equation}%
and the Lagrange multipliers $\lambda (x)$ are determined by using (\ref%
{data})%
\begin{equation}
e^{\lambda (x)}=\frac{Z}{\int d\theta e^{\beta f(\theta )}P_{\text{old}%
}(x,\theta )}\delta (x-x%
{\acute{}}%
)~.
\end{equation}%
The posterior now becomes%
\begin{equation}
P_{\text{new}}(x,\theta )=P_{\text{old}}(x,\theta )\delta (x-x%
{\acute{}}%
)\frac{e^{\beta f(\theta )}}{\zeta (x,\beta )}~,  \label{Posterior-Both}
\end{equation}%
where $\zeta (x,\beta )=\int d\theta e^{\beta f(\theta )}P_{\text{old}%
}(x,\theta ).$

The Lagrange multiplier $\beta $ is determined by first substituting the
posterior into (\ref{moment}),%
\begin{equation}
\int dxd\theta \left[ P_{\text{old}}(x,\theta )\delta (x-x%
{\acute{}}%
)\frac{e^{\beta f(\theta )}}{\zeta (x,\beta )}\right] f(\theta )=F~.
\end{equation}%
Integrating over $x$ yields,%
\begin{equation}
\frac{\int d\theta e^{\beta f(\theta )}P_{\text{old}}(x^{\prime },\theta
)f(\theta )}{\zeta (x^{\prime },\beta )}=F~,
\end{equation}%
where $\zeta (x,\beta )\rightarrow \zeta (x^{\prime },\beta )=\int d\theta
e^{\beta f(\theta )}P_{\text{old}}(x^{\prime },\theta )$. Now $\beta $ can
be determined by%
\begin{equation}
\frac{\partial \ln \zeta (x^{\prime },\beta )}{\partial \beta }=F~.
\label{F}
\end{equation}

The final step is to marginalize the posterior, $P_{\text{new}}(x,\theta )$
over $x$ to get our updated probability,%
\begin{equation}
P_{\text{new}}(\theta )=P_{\text{old}}(x^{\prime },\theta )\frac{e^{\beta
f(\theta )}}{\zeta (x^{\prime },\beta )}
\end{equation}%
Additionally, this result can be rewritten using the product rule as%
\begin{equation}
P_{\text{new}}(\theta )=P_{\text{old}}(\theta )P_{\text{old}}(x^{\prime
}|\theta )\frac{e^{\beta f(\theta )}}{\zeta ^{\prime }(x^{\prime },\beta )}~,
\end{equation}%
where $\zeta ^{\prime }(x^{\prime },\beta )=\int d\theta e^{\beta f(\theta
)}P_{\text{old}}(\theta )P_{\text{old}}(x^{\prime }|\theta ).$ The right
side resembles Bayes theorem, where the term $P_{\text{old}}(x^{\prime
}|\theta )$ is the standard Bayesian likelihood and $P_{\text{old}}(\theta )$
is the prior. The exponential term is a \emph{modification} to these two
terms. Notice when $\beta =0$ (no moment constraint) we recover Bayes' rule.
For $\beta \neq 0$ Bayes' rule is modified by a \textquotedblleft
canonical\textquotedblright\ exponential factor.

It must be noted that MaxEnt has been traditionally used for obtaining a
prior for use in Bayesian statistics. When this is the case, the updating is
sequential. This is not the case here where both types of information are
processed simultaneously. In the sequential updating case, the multiplier $%
\beta $ is chosen so that the posterior $P_{\text{new}}$ only satisfies $%
C_{2}$. In the simultaneous updating case the multiplier $\beta $ is chosen
so that the posterior $P_{\text{new}}$ satisfies both $C_{1}$ and $C_{2}$ or 
$C_{1}\wedge C_{2}$ \cite{GiffinCaticha07}.

\section{The agent example}

\label{Example1}

Let us start with a very simple example: There is a class with $3$ students
sitting in desks next to each other and one professor. The professor
announces that he has a loaded, $3$ sided die and he would like his students
to try to discern the probability of getting a $1$, a $2$ or a $3$. He tells
them that he has created this die in such a way that \emph{on the average},
side $1$ is twice as likely to come up as side $3$. Now he rolls the die
without showing them the results. He announces that he has rolled the die 10
times. Then he writes down how many times a $1$ came up on a piece of paper
and hands it to student A, careful not to let the other students see it. He
proceeds to do this for each of the other students, giving student B the
results of side $2$ and student C the results of side $3.$ What would each
student determine the probabilities of the sides to be? Each needs to
determine the probability of getting \emph{any} particular outcome in one
draw ($\theta _{i}$) given the information.

We summarize the information the following way: there are $3$ agents, A, B
and C. The die is rolled and the counts of each side are represented by, $%
m_{1},m_{2}$ and $m_{3}$ respectively with $n$ representing the total count
so that $n=\sum\nolimits_{i=1}^{3}m_{i}.$ Additionally, we know that on the
average one side, $s_{1}$ is twice as likely to be rolled as $s_{3}.$

The first task is to realize that the correct mathematical model for the
probability of getting a particular side where the information that we have
is the number of sides counted is a multinomial distribution. The
probability of finding $k$ sides in $n$ counts which yields $m_{i}$
instances for the $i^{th}$ side is%
\begin{equation}
P_{\text{old}}(m|\theta ,n)=P_{\text{old}}(m_{1}\ldots m_{k}|\theta
_{1}\ldots \theta _{k},n)=\frac{n!}{m_{1}!\ldots m_{k}!}\theta
_{1}^{m_{1}}\ldots \theta _{k}^{m_{k}}~,  \label{multinomial}
\end{equation}%
where $m=(m_{1},\ldots ,m_{k})$ with $\sum\nolimits_{i=1}^{k}m_{i}=n$, and $%
\theta =(\theta _{1},\ldots ,\theta _{k})$ with $\sum\nolimits_{i=1}^{k}%
\theta _{i}=1$. The general problem is to infer the parameters $\theta $ on
the basis of information about the data, $m^{\prime }.$

Additionally we can include information about the bias of the sides by using
the following general constraint,%
\begin{equation}
\left\langle f(\theta )\right\rangle =F\quad \text{where}\quad f(\theta
)=\sum\nolimits_{i}^{k}f_{i}\theta _{i}~,  \label{cP1}
\end{equation}%
where $f_{i}$ is used to represent the die bias. For our example, on the
average, we will find twice the number of $s_{1}$ as compared to $s_{3}$
thus, \emph{on the average}, the probability of finding one of the sides
will be twice that of the other, $\left\langle \theta _{1}\right\rangle
=2\left\langle \theta _{3}\right\rangle $. In this case, $f_{1}=1,$ $%
f_{3}=-2 $ and $f_{2}=F=0.$

Next we need to write the data (counts) as a constraint which in general is%
\begin{equation}
P(m|n)=\delta (m-m^{\prime })~,  \label{cP2}
\end{equation}%
where $m^{\prime }=\{m_{1}^{\prime },\ldots ,m_{k}^{\prime }\}.$ Finally we
write the appropriate entropy to use,%
\begin{equation}
S[P,P_{\text{old}}]~\text{=}-\sum\limits_{m}\int d\theta P(m,\theta |n)\log 
\frac{P(m,\theta |n)}{P_{\text{old}}(m,\theta |n)}~,  \label{entropyP}
\end{equation}%
where%
\begin{equation}
\sum\limits_{m}=\sum\limits_{m_{1}\ldots m_{k}=0}^{n}\delta
(\sum\nolimits_{i=1}^{k}m_{i}-n)~,
\end{equation}%
and%
\begin{equation}
\int d\theta =\int d\theta _{1}\ldots d\theta _{k}\,\delta \left(
\sum\nolimits_{i=1}^{k}\theta _{i}-1\right) ~,
\end{equation}%
and where $P_{\text{old}}(m,\theta |n)=P_{\text{old}}(\theta |n)P_{\text{old}%
}(m|\theta ,n).$ The prior $P_{\text{old}}(\theta )$ is not important for
our current purpose so for the sake of definiteness we can choose it flat
for our example (there are most likely better choices for priors). We then
maximize this entropy with respect to $P(m,\theta |n)$ subject to
normalization and our constraints which after marginalizing over $m^{\prime
} $ yields,%
\begin{equation}
P(\theta )=P_{\text{old}}(m^{\prime }|\theta ,n)\frac{e^{\beta f(\theta )}}{%
\zeta }~,  \label{posterior P}
\end{equation}%
where%
\begin{equation}
\zeta =\int d\theta \,e^{\beta f(\theta )}P_{\text{old}}(m^{\prime }|\theta
,n)\quad \text{and}\quad F=\frac{\partial \log \zeta }{\partial \beta }~.
\label{zeta P}
\end{equation}%
Notice that if one has no information relating the sides then $\beta =0.$

For our 3 sided die the probability distribution is 
\begin{equation}
P_{\text{e}_{\text{1}}}(\theta _{1},\theta _{2})=\frac{1}{\zeta _{\text{e}}}%
e^{\beta (3\theta _{1}+2\theta _{2}-2)}\theta _{1}^{m_{1}^{\prime }}\theta
_{2}^{m_{2}^{\prime }}(1-\theta _{1}-\theta _{2})^{n-m_{1}^{\prime
}-m_{2}^{\prime }}~.
\end{equation}%
However, each student only has the $m^{\prime }$ that corresponds to their
side. For example, student A has $m_{1}^{\prime }.$ Therefore student A must
marginalize over the unknown information. The result is 
\begin{equation}
\sum_{m_{2}=0}^{n-m_{1}}P_{\text{e}_{\text{1}}}(\theta _{1},\theta _{2})=%
\frac{1}{\zeta _{\text{e}_{\text{1}}}}e^{\beta (3\theta _{1}+2\theta
_{2}-2)}\theta _{1}^{m_{1}^{\prime }}(1-\theta _{1})^{n-m_{1}^{\prime }}~,
\end{equation}%
where $\zeta _{\text{e}_{\text{1}}}$ is the normalization constant. This is
the probability distribution that student A would assign to the die. Since
all of the students will follow the same proper inference method (ME), we
need only look at one of the student's solutions. Notice that all students
or agents agree on some global information, the bias of the die and the
number of total die rolls. However, in general they will determine a
different probability distribution that is dependent on the local
information, in this case the number of rolls of a particular side.

Now imagine that each student's desk is at a vertex of an equilateral
triangle (so that they are equidistant from each other). They notice that
the teacher is looking the other way so they each glance at their neighbor's
paper. Since each of them now have \emph{all} of the information they should
all come up with the same answers.

Next let us create a more complex example by increasing the number of
students. We enlarge the class by adding $k$ students with a professor
rolling a $k$ sided die that is loaded in some given way. The students are
arranged in a lattice structure such as in Figure 1. where there is one
student at each of the vertices. Each student that is not on an edge now has
six neighbors. Thus if they are allowed to 'look' at their nearest
neighbors, the form of the probability distribution that each student would
assign is 
\begin{equation}
P_{\text{e}_{\text{2}}}(\theta _{1}...\theta _{k-1})=\frac{1}{\zeta _{\text{e%
}_{\text{2}}}}e^{\beta f_{k}\left( 1-\sum\limits_{i}^{k-1}\theta
_{i}\right) }(1-\sum\limits_{i}^{7}\theta
_{i})^{n-\sum\limits_{i}^{7}m_{i}}\prod\limits_{i=1}^{7}\theta
_{i}^{m_{i}^{\prime }}\prod\limits_{i=1}^{k-1}e^{\beta f_{i}\theta _{i}}~.
\end{equation}

\begin{figure}[!t]
\centering
 \resizebox{.5\columnwidth}{!}
  {\includegraphics[draft=false]{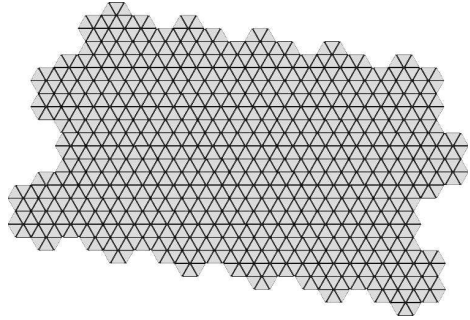}}
  \caption{An example structure that relates agents in a system. Here each vertex is an agent.}
 \label{Fig1:a}
\end{figure}

\section{Conclusions}

\label{Conclusions}

We demonstrated that the ME method can easily lend itself to agent based
modeling. Whether the agents are skin cells, atoms in a lattice, banks in a
network or students in a classroom, the methodology of ME can be applied in
order to model many of these systems. Any system where agents agree on some
global information yet react to local information should be able to be
modeled with this method. It was further shown that the complexity of the
computation can be kept to a minimum since we can marginalize over non-local
data.

By determining what each agent 'thinks' we can predict many properties of
the system. An obvious extension of this work would be to apply decision
theory concepts to the model so as to not only describe how the agents
'think' but what they 'choose' to do as well. This could illustrate how the
agents evolve and could illuminate emergent behavior of the system.

By using the ME method we can include additional information which allows us
to go beyond what Bayes' rule and MaxEnt methods alone could do. Therefore,
we would like to emphasize that anything one can do with Bayesian or MaxEnt
methods, one can now do with ME. Additionally, in ME one now has the ability
to apply additional information that Bayesian or MaxEnt methods could not
process. Further, any work done with Bayesian techniques can be implemented
into the ME method directly through the joint prior.

A currently popular technique is to use entropic concepts on systems.
Whether applying entropy in the thermodynamic sense or from the information
perspective, ME can help here as well. The realization that the ME entropy $%
S_{ME}=\log \zeta +\beta F~$is of the exact same form as the thermodynamic
entropy\footnote{%
The thermodymaical entropy actually has a $-\beta $. Although the ME entropy
has a $+\beta $, the sign is trivial as it is mearly a matter of preference
in our method. We could have substracted the lagrange multipliers instead of
adding them in (\ref{max}).} is of no small consequence. All of the concepts
that thermodynamics utilizes can now also be utilized in models using the ME
methodology, whether it be energy considerations or equilibrium conditions,
etc. In addition, one can get a measure of diversity directly from this
method \cite{GiffinEco07}. To see a detailed method for calculating $\zeta ,$ see \cite{GiffinCaticha07}

\bigskip

\noindent \textbf{Acknowledgements:} We would like to acknowledge many
valuable discussions with A. Caticha.


\newpage

\begin{thebibliography}{9}
\bibitem{Jaynes57} E. T. Jaynes, Phys. Rev. \textbf{106}, 620 and \textbf{108%
}, 171 (1957); R. D. Rosenkrantz (ed.), \emph{E. T. Jaynes: Papers on
Probability, Statistics and Statistical Physics} (Reidel, Dordrecht, 1983);
E. T. Jaynes, \emph{Probability Theory: The Logic of Science} (Cambridge
University Press, Cambridge, 2003).

\bibitem{ShoreJohnson80} J. E. Shore and R. W. Johnson, IEEE Trans. Inf.
Theory \textbf{IT-26}, 26 (1980); IEEE Trans. Inf. Theory \textbf{IT-27}, 26
(1981).

\bibitem{Skilling88} J. Skilling, \textquotedblleft The Axioms of Maximum
Entropy\textquotedblright , \emph{Maximum-Entropy and Bayesian Methods in
Science and Engineering}, G. J. Erickson and C. R. Smith (eds.) (Kluwer,
Dordrecht, 1988).

\bibitem{CatichaGiffin06} A. Caticha and A. Giffin, \textquotedblleft
Updating Probabilities\textquotedblright , \emph{Bayesian Inference and
Maximum Entropy Methods in Science and Engineering}, ed. by Ali
Mohammad-Djafari (ed.), AIP Conf. Proc. \textbf{872}, 31 (2006)
(http://arxiv.org/abs/physics/0608185).

\bibitem{GiffinCaticha07} A. Giffin and A. Caticha, \textquotedblleft
Updating Probabilities with Data and Moments\textquotedblright , \emph{%
Bayesian Inference and Maximum Entropy Methods in Science and Engineering},
ed. by Kevin Knuth, et all, AIP Conf. Proc. \textbf{954}, 74 (2007)
(http://arxiv.org/abs/0708.1593).

\bibitem{Gelman04} A. Gelman, et al., \emph{Bayesian Data Analysis, 2nd
edition} (CRC Press, 2004).

\bibitem{GiffinEcon07} A. Giffin, \textquotedblleft Updating Probabilities
with Data and Moments: An Econometric Example\textquotedblright ,\ presented
at the \emph{3rd Econophysics Colloquium}, Ancona, Italy, 2007  (http://arxiv.org/abs/0710.2912).

\bibitem{GiffinEco07} A. Giffin, \textquotedblleft Infering Diversity: Life
after Shannon\textquotedblright ,\ presented at the \emph{7th International
Conference on Complex Systems}, Boston, 2007 (http://arxiv.org/abs/0709.4079).
\end{thebibliography}
\end{document}